\documentstyle[12pt,a4,epsf]{article}

\def\mylimit#1{\mathrel{\mathop{\kern0pt\longrightarrow}\limits_{#1}}}

\makeatletter
\@addtoreset{equation}{section}
\makeatother

\newcommand{\VEV}[1]{\left\langle #1 \right\rangle}

\newcommand{\nn}{\nonumber}

\newcommand{\bequ}{\begin{equation}}
\newcommand{\eequ}{\end{equation}}
\newcommand{\beqn}{\begin{eqnarray}}
\newcommand{\eeqn}{\end{eqnarray}}
\newcommand{\bctr}{\begin{center}}
\newcommand{\ectr}{\end{center}}

\newcommand{\hsp}[1]{\hspace {#1cm}}
\newcommand{\vsp}[1]{\vspace {#1cm}}

\begin{document}
\begin{titlepage}

\begin{flushright}
hep-ph/0304293\\
KUNS-1840\\
\today
\end{flushright}

\vspace{4ex}

\begin{center}
{\large \bf
Flipped $SO(10)$ model
}

\vspace{6ex}

\renewcommand{\thefootnote}{\alph{footnote}}
Nobuhiro Maekawa\footnote
{e-mail: maekawa@gauge.scphys.kyoto-u.ac.jp
}
and 
Toshifumi Yamashita\footnote{
e-mail: yamasita@gauge.scphys.kyoto-u.ac.jp
}

\vspace{4ex}
{\it Department of Physics, Kyoto University, Kyoto 606-8502, Japan}\\
\end{center}

\renewcommand{\thefootnote}{\arabic{footnote}}
\setcounter{footnote}{0}
\vspace{6ex}

\begin{abstract}
We show that as in the flipped $SU(5)$ models, doublet-triplet splitting
is realized by the missing partner mechanism in the flipped $SO(10)$ models. 
The gauge group $SO(10)_F\times U(1)_{V'_F}$ includes $SU(2)_E$ gauge symmetry,
that plays an important role in solving supersymmetric (SUSY) flavor problem
by introducing non-abelian horizontal gauge symmetry and anomalous $U(1)_A$
gauge symmetry.
The gauge group can be broken into the standard model gauge group by VEVs of
only spinor fields, such models may be easier than $E_6$ models to be derived
from the superstring theory.

\end{abstract}

\end{titlepage}

\section{Introduction}
In the previous paper\cite{horizontal}, 
one of the authors shows that the SUSY flavor problem
can be solved in $E_6$ unification by using non-abelian horizontal gauge 
symmetry and 
anomalous $U(1)_A$ gauge symmetry\cite{U(1)},
whose anomaly is cancelled by the Green-Schwarz mechanism\cite{GS},
 even if large neutrino mixing angles are
obtained. It is essential that the fundamental representation ${\bf 27}$
of $E_6$ has two ${\bf \bar 5}$ fields of $SU(5)$. Actually ${\bf 27}$ is
decomposed as
\begin{equation}
{\bf27} \rightarrow \underbrace{[{\bf 10}_{(1,1)} 
                                  +{\bf \bar 5}_{(1,-3)}
                                  +{\bf 1}_{(1,5)}]}_{{\bf 16}_1}
+\underbrace{[{\bf \bar 5}_{(-2,2)}+{\bf 5}_{(-2,-2)}]}_{{\bf 10}_{-2}}
+ \underbrace{[{\bf 1}_{(4,0)}]}_{{\bf 1}_4}
\end{equation}
under $E_6\supset SO(10)\times U(1)_{V'}\supset 
SU(5)\times U(1)_{V'}\times U(1)_{V}$,
where the representation of $SO(10)\times U(1)_{V'}, 
SU(5)\times U(1)_{V'}\times U(1)_{V}$ are explicitly denoted
in the above.
If we introduce three ${\bf 27}$ fields $\Psi_i$ $(i=1,2,3)$ for three
 generation
quarks and leptons, three of six ${\bf \bar 5}$ fields become massive with three
${\bf 5}$ fields after breaking $E_6$ into $SU(5)$, and the remaining fields
($3\times {\bf \bar 5}$) remain massless. In the $6\times 3$ mass matrix for
${\bf \bar 5}$ and ${\bf 5}$ fields, it is natural to expect that the elements
for the third generation field $\Psi_3$ become larger to realize larger Yukawa
couplings than the first and second generation fields $\Psi_1$ and $\Psi_2$.
Therefore, all the three massless modes of ${\bf \bar 5}$ come mainly from the 
first two generation fields $\Psi_1$ and $\Psi_2$. This structure is interesting
because it can explain larger mixing angles of lepton sector than of quark 
sector as discussed in Ref.\cite{BM}. Moreover, if we introduce non-abelian
horizontal symmetry $SU(2)_H$ and take the first two generation fields as 
doublet,
then all three generation ${\bf \bar 5}$ fields have degenerate sfermion masses,
which are very important to suppress flavor changing neutral current (FCNC) 
processes with large neutrino mixing angles as discussed in 
Ref.\cite{horizontal}. 

In the above arguments, $E_6$ gauge group plays an important role. 
Actually, it is essential that a single field includes two ${\bf \bar 5}$ fields
to realize large neutrino mixing angles with suppressing FCNC processes.
However, in order to break $E_6$ into the standard model (SM) gauge group
$SU(3)_C\times SU(2)_L\times U(1)_Y$, adjoint Higgs fields ${\bf 78}$ are
required, which may not be so easily realized in the framework of superstring 
models. To avoid the adjoint Higgs, it is a simple way to adopt non-simple 
group as a unification group. Which kinds of non-simple group do not
spoil the above interesting features?
The answer is simple. In order to satisfy the essential point that two fields
with the same quantum number under the SM gauge group are included in a single
multiplet, $SU(2)_E$, which is a subgroup of $E_6$ group and rotates
$({\bf \bar 5}_{(1,-3)},{\bf \bar 5}_{(-2,2)})$ and 
$({\bf 1}_{(1,5)},{\bf 1}_{(4,0)})$ as doublets, is sufficient. 
Therefore, it is interesting to consider the unification group which include
$SU(2)_E$. The $SU(3)^3\subset E_6$ is in the case and we know that realistic 
$SU(3)^3$ 
model can be straightforwardly constructed\cite{shafi}, in which 
doublet-triplet splitting problem is solved and realistic quark and lepton
mass matrices are obtained including large neutrino mixing angles.
 Therefore, if we 
introduce non-abelian horizontal symmetry in addition to $SU(3)^3$, 
FCNC processes can be naturally suppressed with large neutrino mixing angles.
In this paper, we consider another non-simple gauge group, 
$SO(10)_F\times U(1)_{V'_F}$,
which can include $SU(2)_E$ because of the unusual embedding of 
the SM gauge group.
We show that in this model, doublet-triplet splitting is realized by missing
partner mechanism. The original missing partner mechanism was introduced in
$SU(5)$ unification group\cite{yanagida}, 
but it requires several large dimensional 
representation Higgs fields. To avoid the large dimensional Higgs fields,
flipped $SU(5)$\cite{Flipped} has been considered. 
It is known that the gauge group 
$SU(5)_F\times U(1)_X$ cannot be unified into $SO(10)$ without spoiling 
the missing partner mechanism, but we show that 
$SO(10)_F\times U(1)_{V'_F}\subset E_6$
can embed the flipped $SU(5)$ without spoiling the missing partner 
mechanism. As noted in the above, the flipped $SO(10)$ gauge group includes
$SU(2)_E$, that is important to solve the SUSY flavor problem by introducing
non-abelian horizontal gauge symmetry and anomalous $U(1)_A$ gauge symmetry.

\section{Review of flipped $SU(5)$ model}
We briefly review the flipped $SU(5)$ model and the reason 
why the flipped $SU(5)$ model cannot be embedded in $SO(10)$ GUT.

It is well-known that one family standard model fermions 
$Q({\bf 3},{\bf 2})_{\frac{1}{6}}$, $U^c({\bf \bar 3},{\bf 1})_{-\frac{2}{3}}$,
$D^c({\bf \bar 3},{\bf 1})_{\frac{1}{3}}$, $L({\bf 1},{\bf 2})_{-\frac{1}{2}}$,
and $E^c({\bf 1},{\bf 1})_1$ plus the right-handed neutrino 
$N^c({\bf 1},{\bf 1})_0$ under the SM gauge group
$SU(3)_C\times SU(2)_L\times U(1)_Y$ are unified into an $SO(10)$-spinorial
${\bf 16}$ superfield:
\begin{equation}
\Psi({\bf 16})\rightarrow {\bf 10}_\Psi({\bf 10}_1)+
{\bf \bar 5}_\Psi({\bf \bar 5}_{-3})+{\bf 1}_\Psi({\bf 1}_5),
\end{equation}
where the decomposition is specified into $SU(5)\times U(1)_V$.
The matter content of the flipped $SU(5)$ models can be obtained from the
corresponding assignment of the standard $SU(5)$ GUT model by means of the 
``flipping" $U^c\leftrightarrow D^c$, $N^c\leftrightarrow E^c$:
\begin{eqnarray}
  {\bf 10}_\Psi&=&(Q,D^c,N^c) \nn \\
{\bf \bar 5}_\Psi&=&(U^c,L) \\
{\bf 1}_\Psi&=&E^c. \nn
\end{eqnarray}
It is important that if ${\bf 10}_1$ representation Higgs ${\bf 10}_C$ 
is introduced,
$SU(5)\times U(1)_X$ can be broken into the standard model gauge
group $SU(3)_C\times SU(2)_L\times U(1)_Y$ by the vacuum expectation value
(VEV) of the component of $N^c$.
Here, the hypercharge operator is written
\begin{equation}
Y=\frac{1}{5}(X-Y'),
\end{equation}
where $Y'$ is the generator of $SU(5)_F$ which commutes with 
$SU(3)_C\times SU(2)_L$. Then the $SO(10)$-vectorial ${\bf 10}$
superfield decomposed as
\begin{equation}
H({\bf 10})\rightarrow {\bf 5}_H({\bf 5}_{-2})+{\bf \bar 5}_H({\bf \bar 5}_{2})
\end{equation}
includes the SM doublet Higgs $H_d=L'$ and $H_u=\bar L'$ as
\begin{eqnarray}
  {\bf 5}_H&=&(\bar D^{c\prime},L') \nn \\
{\bf \bar 5}_H&=&(D^{c\prime}, \bar L'),
\end{eqnarray}
where $D^{c\prime}$ and $L'$ have the same quantum number of SM gauge group
as $D^c$ and $L$, respectively.
If we introduce interactions in the superpotential as
\begin{equation}
W_{MP}={\bf 10}_C{\bf 10}_C {\bf 5}_H
+{\bf \overline{10}}_{\bar C}{\bf \overline{10}}_{\bar C}{\bf \bar 5}_{H},
\label{MP}
\end{equation}
only the triplet Higgs $\bar D^{c\prime}$ and $D^{c\prime}$ can be 
superheavy  with
$D^c$ in ${\bf 10}_C$ and $\bar D^c$ in ${\bf \overline{10}}_{\bar C}$,
respectively, by developing the VEVs of ${\bf 10}_{C}$ and
${\bf \overline{10}}_{\bar C}$, but the doublet Higgs $L'$ and 
$\bar L'$ have no partner and remain massless.
This is essential of the missing partner 
mechanism in the flipped $SU(5)$ model.

Unfortunately, this missing partner mechanism in the flipped $SU(5)$ model 
cannot be extended to $SO(10)$ unification. In $SO(10)$ 
unification the interactions (\ref{MP}) are included in the $SO(10)$ symmetric
interactions $C({\bf 16})C({\bf 16})H({\bf 10})$ and 
$\bar C({\bf \overline{16}})\bar C({\bf \overline{16}}) H({\bf 10})$, 
which include also
\begin{equation}
 {\bf 10}_C{\bf \bar 5}_C {\bf \bar 5}_H
+{\bf \overline{10}}_{\bar C}{\bf 5}_{\bar C}{\bf 5}_{H}.
\label{BMP}
\end{equation}
Through these interactions, the doublet Higgs $(\bar L')_H$ and $(L')_H$ become
superheavy with ${L}_C$ and $(L^*)_{\bar C}$, respectively, 
by developing
the VEVs of ${\bf 10}_{\bar C}$ and
${\bf \overline{10}}_{\bar C}$. (In this paper, $X^*$ is a component of
${\bf \overline{16}}$ of $SO(10)$ and denotes 
the complex conjugate representation of $X$ which is a component of
${\bf 16}$ of $SO(10)$. ) 
Therefore, doublet-triplet splitting is spoiled
by this extension.\footnote{Of course, if we neglect the component fields 
${\bf \bar 5}_C$ and ${\bf 5}_{\bar C}$ by hand, such extension becomes possible
\cite{valle}.}

In the next section, we show that the missing partner mechanism of the flipped
$SU(5)$ model can be embedded in $SO(10)_F\times U(1)_{V'_F}$ unification 
group.

\section{Flipped $SO(10)$ model}
As noted in the introduction, 
${\bf 27}$ of $E_6$
is decomposed as
\begin{equation}
{\bf27} \rightarrow \underbrace{[{\bf 10}_{(1,1)} 
                                  +{\bf \bar 5}_{(1,-3)}
                                  +{\bf 1}_{(1,5)}]}_{{\bf 16}_1}
+\underbrace{[{\bf \bar 5}_{(-2,2)}+{\bf 5}_{(-2,-2)}]}_{{\bf 10}_{-2}}
+ \underbrace{[{\bf 1}_{(4,0)}]}_{{\bf 1}_4}
\end{equation}
under $E_6\supset SO(10)\times U(1)_{V'}\supset 
SU(5)\times U(1)_{V'}\times U(1)_{V}$.
There are two ways to embed the flipped $SU(5)$ matters
${\bf 10}_\Psi=(Q,D^c,N^c)$, ${\bf \bar 5}_\Psi=(U^c,L)$ and 
${\bf 1}_\Psi=E^c$ in the above decomposition of ${\bf 27}$ of $E_6$
into $SO(10)\times U(1)_{V'}$. As discussed in the previous section,
the usual embedding $SU(5)_F\times U(1)_X$ in $SO(10)$,
\begin{equation}
\underbrace{[{\bf 10}_\Psi+{\bf \bar 5}_\Psi+{\bf 1}_\Psi]}_{{\bf 16}_1}
+\underbrace{[{\bf \bar 5}_H+{\bf 5}_H]}_{{\bf 10}_{-2}}
+ \underbrace{[{\bf 1}_S]}_{{\bf 1}_4},
\end{equation}
where ${\bf 5}_H=(\bar D^{c\prime},L')$, 
${\bf \bar 5}_H=(D^{c\prime}, \bar L')$ and
${\bf 1}_S$ is singlet under $SU(5)_F\times U(1)_X$,
 spoils the 
missing partner mechanism. The other embedding can be obtained by
means of the ``flipping"
${\bf \bar 5}_\Psi \leftrightarrow {\bf \bar 5}_H$ and
${\bf 1}_\Psi \leftrightarrow {\bf 1}_S$:
\begin{equation}
\underbrace{[{\bf 10}_\Psi+{\bf \bar 5}_H+{\bf 1}_S]}_{{\bf 16}_1}
+\underbrace{[{\bf \bar 5}_\Psi+{\bf 5}_H]}_{{\bf 10}_{-2}}
+ \underbrace{[{\bf 1}_\Psi]}_{{\bf 1}_4}.
\end{equation}
In this embedding, if ${\bf 1}_S$ component of ${\bf 16}_1$ field have 
non-vanishing
VEV, $SO(10)_F\times U(1)_{V'_F}$ is broken into $SU(5)_F\times U(1)_X$.
Here, the operator $X$ is obtained as
\begin{equation}
X=\frac{1}{4}(5V'_F-V_F),
\end{equation}
where $V_F$ is the generator of $SO(10)_F$ which commute with
$SU(5)_F$. The hypercharge operator is
\begin{equation}
Y=\frac{1}{5}(X-Y')=\frac{1}{20}(5V'_F-V_F-4Y').
\end{equation}
Note that the each $SU(2)_E$ doublet $(D^{c\prime},D^c)$, $(L',L)$ and 
$(N^c,S)$,
which has the same quantum number of SM gauge group, 
is included into a single multiplet ${\bf 16}_1$, ${\bf 10}_{-2}$ and 
${\bf 16}_1$, respectively. 
This means that $SU(2)_E$ is embedded in $SO(10)_F$.

We introduce two pairs of Higgs fields 
$[\Phi({\bf 16}_1),\bar \Phi({\bf \overline{16}}_{-1})]$ and
$[C({\bf 16}_1),\bar C({\bf \overline{16}}_{-1})]$ to break 
$SO(10)_F\times U(1)_{V'_F}$ into the SM gauge group. Supposing that
the VEVs 
$|\VEV{\Phi}|=|\VEV{\bar \Phi}|$ breaks $SO(10)_F\times U(1)_{V'_F}$
into $SU(5)_F\times U(1)_X$, the components ${\bf 10}_\Phi$ and 
${\bf\overline{10}}_{\bar \Phi}$ are absorbed by the Higgs mechanism.
The VEVs $|\VEV{C}|=|\VEV{\bar C}|$ break $SU(5)_F\times U(1)_X$ into
the SM gauge group, and the components $Q$ and $N^c$ are absorbed by
the Higgs mechanism.
All the remaining components ${\bf \bar 5}_\Phi$, ${\bf 5}_{\bar \Phi}$, 
${\bf \bar 5}_C$, ${\bf 5}_{\bar C}$, $(D^c)_C$ and $(D^{c*})_{\bar C}$
must be massive except a pair of doublets. For example,
through the interactions in the superpotential,
\begin{equation}
W_{SO(10)}=\bar \Phi\bar \Phi C C+\bar C\bar C\Phi\Phi,
\label{SO}
\end{equation}
which include the interactions (\ref{MP}) after developing the VEVs
$|\VEV{\Phi}|=|\VEV{\bar \Phi}|$, 
pairs $[(D^{c\prime*})_{\bar \Phi}, (D^c)_C]$ and
$[(D^{c\prime})_{\Phi}, (D^{c*})_{\bar C}]$ become massive. If we introduce 
the mass term for $C$ and $\bar C$, then only
$(\bar L')_\Phi$ and $(\bar L^{\prime *})_{\bar \Phi}$ remain massless, namely,
doublet-triplet splitting is realized.
There are several interactions which unstabilize the doublet-triplet
splitting. For example, the terms $\bar\Phi\Phi F(\bar CC,\bar \Phi\Phi)$
give directly the doublet Higgs mass, so they must be forbidden.
(We will return to this subject lator in a concrete model.)

We assume that three generation matter fields
$\Psi_i({\bf 27})={\bf 16}_{\Psi_i}+{\bf 10}_{\Psi_i}+{\bf 1}_{\Psi_i}$
$(i=1,2,3)$ respect $E_6$ symmetry. It is an easy way to guarantee the 
cancellation of gauge anomaly. Among the three generation matter fields
$\Psi_i$, there are six fields which have the same quantum number under
the SM gauge group as $(D^c,L)$. Only three linear combinations of these fields
become quarks and leptons, and other modes become superheavy with the 
three $(\bar D^{c\prime},\bar L')$ fields through the interactions
${\bf 16}_{\Psi_i}{\bf 10}_{\Psi_j}\Phi$ and 
${\bf 16}_{\Psi_i}{\bf 10}_{\Psi_j}C$ by developing the VEVs of $\Phi$ and $C$.
It is interesting that up-type Yukawa coupling can be obtained from
the renormalizable interactions ${\bf 16}_{\Psi_i}{\bf 10}_{\Psi_j}\Phi$, 
because $O(1)$ top Yukawa coupling can be naturally realized.
But Yukawa couplings of down quark sector and of charged lepton sector
are obtained from the higher dimensional interactions
${\bf 16}_{\Psi_i}{\bf 16}_{\Psi_j}\bar C\bar \Phi$ and 
${\bf 10}_{\Psi_i}{\bf 1}_{\Psi_j}\bar C\bar \Phi$, respectively.
Because we have six singlets $N^c_i$ and $S_i$ in the matter sector, 
the mass matrix for right-handed neutrino becomes $6\times 6$ matrix which
are obtained from the interactions
$\Psi_i\Psi_j\bar\Phi\bar\Phi$, $\Psi_i\Psi_j\bar\Phi\bar C$ and
$\Psi_i\Psi_j\bar C\bar C$.
Yukawa couplings of Dirac neutrino sector
are obtained from the interactions ${\bf 16}_{\Psi_i}{\bf 10}_{\Psi_j}\Phi$.
Therefore, the mass terms of all quarks and leptons can be obtained
in this scenario.

Unfortunately, as in the flipped $SU(5)$ model, 
this missing partner mechanism in the 
flipped $SO(10)$ model cannot be extended to $E_6$ unification. 
In $E_6$ 
unification the interactions (\ref{SO}) are included in the $E_6$ symmetric
interactions 
$\Phi({\bf 27})\Phi ({\bf 27})\bar C({\bf \overline{27}})
\bar C({\bf \overline{27}})$ and $\bar \Phi({\bf \overline{27}})
\bar \Phi ({\bf \overline{27}})C({\bf {27}})
C({\bf {27}})$, which include also
${\bf 16}_\Phi{\bf 10}_\Phi{\bf 10}_{\bar C}{\bf \overline{16}}_{\bar C}$
and ${\bf 16}_C{\bf 10}_C{\bf 10}_{\bar \Phi}{\bf \overline{16}}_{\bar \Phi}$
of $SO(10)_F$.
After developing the VEVs $|\VEV{\Phi}|=|\VEV{\bar \Phi}|$, these interactions
give
${\bf 5}_\Phi{\bf 5}_{\bar C}{\bf \overline{10}}_{\bar C}$
and
${\bf \bar 5}_{\bar \Phi}{\bf \bar 5}_{C}{\bf {10}}_{C}$
of $SU(5)_F$,
which give mass terms to doubet Higgs by taking non-vanishing VEVs 
$|\VEV{C}|=|\VEV{\bar C}|$.
Therefore, doublet-triplet splitting is spoiled in this extension.

\section{Flipped $SO(10)$ model with anomalous $U(1)_A$}
It is important to find a concrete flipped $SO(10)$ model in which
doublet-triplet splitting is realized with generic interactions and
to examine whether the realistic quark and lepton mass matrices are 
realized or not. 
In a series of papers\cite{horizontal,BM,shafi,TGUT,MY,Unif},
we have pointed out that anomalous $U(1)_A$ symmetry
 plays an important
role in solving various problems in SUSY grand unified theory (GUT)
with generic interactions. This is mainly because the SUSY zero 
mechanism (holomorphic zero)\footnote{Note that if the total charge
of an operator is negative, the $U(1)_A$ invariance and 
analytic property of the superpotential forbids 
the existence of the operator in the superpotential, 
since the Froggatt-Nielsen
\cite{FN} field $\Theta$ with negative 
charge cannot compensate for the negative total charge of the operator 
(the SUSY zero mechanism).}
can control various terms which must be forbidden.

In this section, we present a concrete flipped
$SO(10)$ model with generic interaction by 
introducing anomalous $U(1)_A$ symmetry.
\subsection{Higgs sector}
The Higgs contents are listed in Table I.
\vspace{3mm}
\begin{center}
Table I. The typical values of anomalous $U(1)_A$ charges are listed.
$\pm$ is $Z_2$-parity and $i=1, 2$.

\begin{tabular}{|c|c|c|} 
\hline
                  &   non-vanishing VEV  & vanishing VEV \\
\hline 
${\bf 16}_1$
                  &   $\Phi(\phi=0,-)$\  $C(c=-2,+)$ 
                  &  $\Phi_i'(\phi_i'=5,-)$  \\
${\bf \overline{16}}_{-1}$
                  & $\bar \Phi(\bar \phi=-1,-)$ \  $\bar C(\bar c=-2,+)$ &
                  $\bar \Phi_i'(\bar\phi_i'=4,-)$ \\
{\bf 1}           &   $\Theta(\theta=-1,+)$\ $\bar Z_i(\bar z_i=-1,+)$ 
                    \ $Z(z=-4,-)$  & $S'(s'=8,+)$
\\
\hline
\end{tabular}

\vspace{5mm}
\end{center}
Following the general discussion on the determination of VEVs of the
models with anomalous $U(1)_A$ charges, only the negatively charged
fields can have non-vanishing VEVs\cite{BM,TGUT,MY,Unif}.
The scale of these VEVs are determined by the anomalous $U(1)_A$ charges as
\begin{equation}
\VEV{\bar\Phi\Phi}\sim \lambda^{-(\phi+\bar\phi)},\ 
\VEV{\bar CC}\sim \lambda^{-(c+\bar c)},
\end{equation}
where $\lambda$ is the ratio of the VEV of Froggatt-Nielsen field $\Theta$,
which is essentially determined by the Fayet-Illiopoulos $D$-term parameter,
to the cutoff $\Lambda$. In this paper, we take $\lambda$ as around the Cabbibo
angle $\sin \theta_W\sim 0.22$.
If the ${\bf 1}_{(1,5)}$ component of $\Phi$ and the ${\bf 1}_{(-1,-5)}$ 
component of $\bar \Phi$ have non-vanishing VEVs, 
$SO(10)_F\times U(1)_{V'_F}$ is broken into $SU(5)_F\times U(1)_X$. 
The ${\bf 10}_{(1,1)}$ of $\Phi$ and ${\bf \overline{10}}_{(-1,-1)}$ of 
$\bar \Phi$ are absorbed by the Higgs mechanism at that time. 
Moreover, if the ${\bf 10}_{(1,1)}$ component of $C$
and the ${\bf \overline{10}}_{(-1,-1)}$ component of $\bar C$
have non-vanishing VEVs, $SU(5)_F\times U(1)_X$ is broken into the SM gauge 
group. Then the $Q$ component of ${\bf 10}_{(1,1)}$ of $C$ and 
the $\bar Q$ component of ${\bf \overline{10}}_{(-1,-1)}$ of $\bar C$ are 
absorbed by the Higgs mechanism. 
Therefore, the remaining negatively charged fields except
singlets under the SM gauge group
are the ${\bf \bar 5}_{(1,-3)}$ components of $\Phi$ and $C$, the 
$D^c$ component of $C$, and the mirror components of $\bar \Phi$ and $\bar C$. 
Among these negatively charged fields, no mass term appears because
of the SUSY zero (holomorphic zero) mechanism. In order to make them massive, 
we have to take account of the positively charged fields $\Phi'_i$ and 
$\bar\Phi'_i$.
Note that in a ${\bf 16}_1$ field, there are two colored Higgs 
$D^c$ and $D^{c'}$ because of $SU(2)_E$ symmetry, but only one doublet
$\bar L'$. Therefore, the colored Higgs mass matrix becomes $7\times 7$ matrix
$M_T$ which is given by
\begin{equation}
\bordermatrix{
 \bar D^c\backslash D^c & {\bf 10}_C &{\bf \bar 5}_C& {\bf \bar 5}_\Phi &
                          {\bf 10}_{\Phi'_1} & {\bf 10}_{\Phi'_2} &
                          {\bf \bar 5}_{\Phi'_1} & {\bf \bar 5}_{\Phi'_2} \cr
{\bf \overline{10}}_{\bar C} &0& 0 & 0 & 0 & 0 & 
                             \lambda^{\bar c+\phi'_1-\Delta} &
                             \lambda^{\bar c+\phi'_2-\Delta} \cr
{\bf 5}_{\bar C}   &0& 0 & 0 & \lambda^{\bar c+\phi'_1+\Delta} &
                             \lambda^{\bar c+\phi'_2+\Delta} &
                             0 & 0 \cr
{\bf 5}_{\bar \Phi} & 0 & 0 & 0 & 0 & 0 & \lambda^{\bar\phi+\phi'_1} &
                                        \lambda^{\bar\phi+\phi'_2} \cr
{\bf \overline{10}}_{\bar\Phi'_1} & 0 & 
                \lambda^{\bar\phi'_1+c-\Delta} & 0 &
                \lambda^{\bar\phi'_1+\phi'_1} & \lambda^{\bar\phi'_1+\phi'_2}&
                \lambda^{\bar\phi'_1+\phi'_1-\Delta} &
                \lambda^{\bar\phi'_1+\phi'_1-\Delta} \cr
{\bf \overline{10}}_{\bar\Phi'_2} & 0 & 
                \lambda^{\bar\phi'_2+c-\Delta} & 0 &
                \lambda^{\bar\phi'_2+\phi'_1} & \lambda^{\bar\phi'_2+\phi'_2}&
                \lambda^{\bar\phi'_2+\phi'_1-\Delta} &
                \lambda^{\bar\phi'_2+\phi'_1-\Delta} \cr
{\bf 5}_{\bar\Phi'_1} & \lambda^{\bar\phi'_1+c+\Delta} & 0 & 
                    \lambda^{\bar\phi'_1+\phi} &
                   \lambda^{\bar\phi'_1+\phi'_1+\Delta} &
                \lambda^{\bar\phi'_1+\phi'_2+\Delta} &
                \lambda^{\bar\phi'_1+\phi'_1} & 
                \lambda^{\bar\phi'_1+\phi'_2} \cr
{\bf 5}_{\bar\Phi'_2} & \lambda^{\bar\phi'_2+c+\Delta} & 0 & 
                    \lambda^{\bar\phi'_2+\phi} &
                   \lambda^{\bar\phi'_2+\phi'_1+\Delta} &
                \lambda^{\bar\phi'_2+\phi'_2+\Delta} &
                \lambda^{\bar\phi'_2+\phi'_1} & 
                \lambda^{\bar\phi'_2+\phi'_2} \cr
}, 
\label{Tmass10}
\end{equation}
where $\Delta\equiv \frac{1}{2}(\bar \phi-\phi-\bar c+ c)$.
The rank becomes seven for the charge assignment in Table I.
On the other hand, the mass matrix for doublet Higgs becomes $4\times 4$ matrix.
The charges in Table I lead to 
\begin{equation}
M_D=\bordermatrix{
 (\bar L^{\prime*})\backslash \bar L' & C & \Phi & \Phi'_1 &\Phi'_2 \cr
\bar C &0& 0 & 0 & 0 \cr
\bar \Phi   &0& 0 & \lambda^{\phi'_1+\bar \phi} 
                  & \lambda^{\phi'_2+\bar \phi} \cr
\bar \Phi'_1&0& \lambda^{\phi+\bar\phi'_1} & \lambda^{\phi'_1+\bar \phi'_1} 
          & \lambda^{\phi'_2+\bar \phi'_1} \cr
\bar \Phi'_2&0& \lambda^{\phi+\bar\phi'_2} & \lambda^{\phi'_1+\bar \phi'_2} 
          & \lambda^{\phi'_2+\bar \phi'_2} \cr
}.
\label{Dmass10}
\end{equation}
It is obvious that the rank is reduced to three, and therefore one pair of
doublet Higgs appears in this model. The massless modes are written
\begin{eqnarray}
H_u&=&(\bar L')_C,\label{Hu}\\
H_d&=& (\bar L^{\prime *})_{\bar C}
\label{Hd}
\end{eqnarray}
where $H_u$ and $H_d$ are the doublet Higgs for up-quark sector and for
down-quark sector, respectively.

\subsection{Quark and lepton sector}
In this subsection, we use the standard definition of 
${\bf \bar 5}\equiv(D^c, L)$ field.
If we introduce three generation matter fields 
$\Psi_i({\bf 27})={\bf 16}_{\Psi_i}+{\bf 10}_{\Psi_i}+{\bf 1}_{\Psi_i}$
$(i=1,2,3)$ with their charges $(\psi_1,\psi_2,\psi_3)=(4,3,1)$
in addition to the Higgs sector in Table I, 
the massless modes of ${\bf \bar 5}$ fields, where we have 
used the usual definition for ${\bf \bar 5}$, become
\begin{eqnarray}
{\bf \bar 5}_1&=&{\bf \bar 5'}_{\Psi_1}+\lambda^3{\bf \bar 5'}_{\Psi_3}
                +\lambda^{1.5}{\bf \bar 5}_{\Psi_2}
                +\lambda^{3.5}{\bf \bar 5}_{\Psi_3} \nn \\
{\bf \bar 5}_2&=&{\bf \bar 5}_{\Psi_1}+\lambda^{2.5}{\bf \bar 5'}_{\Psi_3}
                +\lambda^{1}{\bf \bar 5}_{\Psi_2}
                +\lambda^{3}{\bf \bar 5}_{\Psi_3} \label{mixing} \\
{\bf \bar 5}_3&=&{\bf \bar 5'}_{\Psi_2}+\lambda^2{\bf \bar 5'}_{\Psi_3}
                +\lambda^{0.5}{\bf \bar 5}_{\Psi_2}
                +\lambda^{2.5}{\bf \bar 5}_{\Psi_3},\nn
\end{eqnarray}
where ${\bf \bar 5'}\equiv(D^{c\prime}, L')$ 
and we fix the three bases of the massless modes
$({\bf \bar 5}_1,{\bf \bar 5}_2,{\bf \bar 5}_3)$ to 
$({\bf \bar 5'}_{\Psi_1},{\bf \bar 5}_{\Psi_1},{\bf \bar 5'}_{\Psi_2})$.
These are obtained from the mass matrix of three ${\bf 5}$ fields and
six ${\bf \bar 5}$ fields which are given from the interactions
$\Psi_i\Psi_j\Phi Z$ and $\Psi_i\Psi_jC$ by developing the VEVs of $\Phi$,
$C$ and $Z$. Then we can estimate the Yukawa couplings of quarks and leptons.

The Yukawa couplings of up quark sector
are obtained as
\begin{equation}
Y_u=\bordermatrix{ & U^c_{\Psi_1} & U^c_{\Psi_2} & U^c_{\Psi_3} \cr
                 Q_{\Psi_1} & \lambda^6 & \lambda^5 & \lambda^3 \cr
                 Q_{\Psi_2} & \lambda^5 & \lambda^4 & \lambda^2 \cr
                 Q_{\Psi_3} & \lambda^3 & \lambda^2 & 1 \cr}
\end{equation}
from the interactions 
$\lambda^{\psi_i+\psi_j+c}{\bf 16}_{\Psi_i}{\bf 10}_{\Psi_j}C$.
The Yukawa couplings of down quark sector and of charged lepton sector
are given as
\begin{equation}
Y_d^T(\sim Y_e)=\bordermatrix{ & Q_{\Psi_1}(E^c_{\Psi_1}) 
                                 & Q_{\Psi_2}(E^c_{\Psi_2}) 
                                 & Q_{\Psi_3}(E^c_{\Psi_3}) \cr
                {\bf \bar 5}_1 &  \lambda^6 & \lambda^5 & \lambda^3 \cr
                 {\bf \bar 5}_2 &  \lambda^{5.5} & \lambda^{4.5} & 0 \cr
                  {\bf \bar 5}_3 & \lambda^5 & \lambda^4 & \lambda^2 \cr}
\end{equation}
 from the higher dimensional interactions
$\lambda^{\psi_i+\psi_j+2\bar c}
{\bf 16}_{\Psi_i}{\bf 16}_{\Psi_j}\bar C\bar C$ and 
$\lambda^{\psi_i+\psi_j+2\bar c}
{\bf 10}_{\Psi_i}{\bf 1}_{\Psi_j}\bar C\bar C$, respectively.
Note that only ${\bf \bar 5'}$ fields can have non-vanishing Yukawa couplings
through the interactions. This is because the interactions 
${\bf 16}_{\Psi_i}{\bf 16}_{\Psi_j}\bar C\bar \Phi$ and
${\bf 10}_{\Psi_i}{\bf 1}_{\Psi_j}\bar C\bar \Phi$ are forbidden by 
$Z_2$-parity. 
The above mass matrices give almost good values for masses and 
mixings for quark sector and charged lepton sector.

The Yukawa couplings for the Dirac neutrino are given as
\begin{equation}
Y_{n_D}=\bordermatrix{& N^c_{\Psi_1} &N^c_{\Psi_2} &N^c_{\Psi_3} &
                      S_{\Psi_1} &S_{\Psi_2} &S_{\Psi_3} \cr
               {\bf \bar 5}_1& \lambda^{6.5} & \lambda^{5.5} & \lambda^{3.5} &
                      \lambda^6 & \lambda^5 & \lambda^3 \cr
                {\bf \bar 5}_2&  \lambda^{6} & \lambda^{5} & \lambda^{3} &
                  \lambda^{5.5} & \lambda^{4.5} & \lambda^{2.5} \cr
                {\bf \bar 5}_3&  \lambda^{5.5} & \lambda^{4.5} & \lambda^{2.5} &
                  \lambda^5 & \lambda^4 & \lambda^2 \cr}
\end{equation}
through the interactions 
$\lambda^{\psi_i+\psi_j+c}{\bf 10}_{\Psi_i}{\bf 16}_{\Psi_j}C$.
The vanishing component is caused by SUSY zero (holomorphic zero).
The right-handed neutrino mass matrix becomes
\begin{eqnarray}
M_{nR}&=&\bordermatrix{ & N^c_{\Psi_1} &N^c_{\Psi_2} &N^c_{\Psi_3} &\hsp{-0}&
                      S_{\Psi_1} &S_{\Psi_2} &S_{\Psi_3} \cr
         N^c_{\Psi_1}&  \lambda^8 & \lambda^7 & \lambda^5 
                     &\hsp{-0}&  \lambda^{7.5} &\lambda^{6.5} & 0\cr
         N^c_{\Psi_2}&  \lambda^7 & \lambda^6 & \lambda^4 
                     &\hsp{-0}&  \lambda^{6.5} & 0 & 0 \cr
         N^c_{\Psi_3}&  \lambda^5 & \lambda^4 & 0 
                     &\hsp{-0}&  0 & 0 & 0 \cr \vsp{-0.1}\cr
          S_{\Psi_1} &  \lambda^{7.5} & \lambda^{6.5} & 0 
                     &\hsp{-0}&\lambda^{7} & \lambda^{6} & \lambda^{4} \cr
        S_{\Psi_2} &  \lambda^{6.5} & 0 & 0 
                   &\hsp{-0}& \lambda^{6} & \lambda^{5} & \lambda^{3} \cr
         S_{\Psi_3} &  0 & 0 &  0 
                    &\hsp{-0}&  \lambda^{4} & \lambda^{3} & \lambda \cr}\Lambda
\end{eqnarray}
through the interactions ${\bf 16}_{\Psi_i}{\bf 16}_{\Psi_j}\bar C\bar C$,
${\bf 16}_{\Psi_i}{\bf 16}_{\Psi_j}\bar C\bar \Phi Z$
${\bf 16}_{\Psi_i}{\bf 16}_{\Psi_j}\bar \Phi\bar \Phi$.
Here vanishing components are caused by the SUSY zero (holomorphic zero) 
mechanism.
Then the neutrino mass matrix is given by
\begin{equation}
M_\nu=Y_{n_D}M_{nR}^{-1}Y_{n_D}^T\VEV{H_u}^2\eta^2\sim \lambda^3
    \left(\matrix{\lambda^2 & \lambda^{1.5} & \lambda \cr
                  \lambda^{1.5} & \lambda & \lambda^{0.5} \cr
                  \lambda & \lambda^{0.5} & 1 \cr}\right)
\frac{\VEV{H_u}^2\eta^2}{\Lambda},
\end{equation}
where $\eta$ is a renormalization factor. This gives bi-large neutrino 
mixings but to realize the mass scale for the neutrino, we have to take
the cutoff $\Lambda\sim 10^{13}$ GeV  if we take 
$\VEV{H_u}\eta\sim 200$ GeV. Such a small cutoff scale leads to too short
neucleon life-time via dimension six operators.
Therefore, the charge assignment in Table I looks unrealistic.

However, because the neutrino scale is determined by the anomalous $U(1)_A$
charges as
\begin{eqnarray}
M_\nu&\sim &\lambda^{-5-l}
    \left(\matrix{\lambda^2 & \lambda^{1.5} & \lambda \cr
                  \lambda^{1.5} & \lambda & \lambda^{0.5} \cr
                  \lambda & \lambda^{0.5} & 1 \cr}\right)
\frac{\VEV{H_u}^2\eta^2}{\Lambda}, \\
l&=&-2c+\bar c-10,
\end{eqnarray}
there may be other realistic models with other charge assignments.
To obtain larger value of $l$, smaller $c$ and/or larger $\bar c$ is needed.
Because $C$ includes $H_u$, the charge $c$ is determined as $c=-2\psi_3=-2n$ 
so that the top Yukawa coupling becomes $O(1)$. Here we take $\psi_i=\delta_i+n$
[$(\delta_1,\delta_2,\delta_3)=(3,2,0)$] to obtain realistic 
Cabbibo-Kobayashi-Maskawa matrix. To realize bi-large neutrino mixings (i.e.,
${\bf \bar 5}$ fields in Eq. (\ref{mixing})), we must take
\begin{equation}
r\equiv \frac{1}{2}[(c-\bar c)-(\phi-\bar \phi)]\sim -\frac{1}{2}.
\end{equation}
Once we fix the mixing structure of ${\bf \bar 5}$ fields, the Yukawa 
couplings for down quarks are proportional to
$\lambda^{\psi_i+\psi_j+2\bar c}\VEV{\bar C}\sim \lambda^{\delta_i+\delta_j
+\frac{3}{2}(\bar c-c)}$. Therefore, roughly speaking, 
$\tan\beta\equiv\frac{\VEV{H_u}}{\VEV{H_d}}$ is proportional to
$\lambda^{\frac{3}{2}(\bar c-c)}$. Then, smaller $c$ and/or larger $\bar c$
lead to smaller $\tan\beta$. For fixed $\tan\beta$, smaller $c$ and $\bar c$
lead to larger $l$. However, unless the condition 
\begin{equation}
c-2\bar c\leq 2
\end{equation}
is satisfied, $(Y_d)_{33}$ vanishes by the SUSY zero mechanism.
A compromised charge assignment is
$(\phi,\bar\phi,c,\bar c,\phi'_i,\bar \phi'_i,\bar z_i, z,s')=
 (-1,-1,-4,-3,8,8,-1,-6,12)$.
Then $l$ becomes $-5$, so the cutoff scale can be larger than the 
$10^{15}$ GeV. Actually, the running gauge couplings of $SU(3)_C$ and $SU(2)_L$,
which should meet at the cutoff scale in this flipped $SO(10)$ scenario,
meet around the scale in this charge assignment. 
And the Yukawa coupling of bottom quark becomes $\lambda^{3.5}$ which can be 
realistic although the large ambiguity of $O(1)$ coefficients is required.

\section{Summary}
In this paper, we have shown that the missing partner mechanism in 
flipped $SU(5)$ model can
be embedded in flipped $SO(10)$ model whose gauge group is 
$SO(10)_F\times U(1)_{V'_F}\subset E_6$. It is interesting that the gauge 
group includes
$SU(2)_E$, that plays an important role in solving SUSY flavor problem
by the horizontal gauge symmetry and anomalous $U(1)_A$ gauge symmetry.
As a proof of existence of a concrete model, we build a flipped $SO(10)$ 
model by introducing anomalous $U(1)_A$ gauge symmetry. 

\section*{Acknowledgement}
  N.M. is supported in part by Grants-in-Aid for Scientific 
Research from the Ministry of Education, Culture, Sports, Science 
and Technology of Japan.


\begin{thebibliography}{99}
\bibitem{horizontal} N. Maekawa, hep-ph/0212141, to appear in 
                    {\it Phys. Lett.} {\bf B}; hep-ph/0304076.
\bibitem{U(1)}    E.~Witten,  Phys. Lett. {\bf B149}, 351 (1984);
                  M.~Dine, N.~Seiberg and E.~Witten,
                  Nucl. Phys. {\bf B289}, 589 (1987);
                  J.J.~Atick, L.J.~Dixon and A.~Sen,
                  Nucl. Phys. {\bf B292}, 109 (1987);
                  M.~Dine, I.~Ichinose and N.~Seiberg,
                  Nucl. Phys.  {\bf B293}, 253 (1987).
\bibitem{GS}      M.~Green and J.~Schwarz,
                  Phys. Lett. {\bf B149}, 117 (1984).
\bibitem{BM}      M. Bando and N. Maekawa, Prog. Theor. Phys. {\bf 106},
                   1255  (2001);
\bibitem{shafi}    N. Maekawa and Q. Shafi, Prog. Theor. Phys. {\bf 109},
                  279 (2003).
\bibitem{yanagida} A. Masiero, D.V. Nanopoulos, K. Tamvakis and T.Yanagida,
                      Phys. Lett. {\bf 115} (1982) 380;\\
                      B. Grinstein, Nucl. Phys. {\bf B206} (1982) 387.
\bibitem{Flipped} S.M. Barr, Phys. Lett. {\b f B112} (1982) 219:\\
                   I. Antoniadis, J. Ellis, J.S. Hagelin and D.V. Nanopoulos,
                     Phys. Lett. {\bf 194B} (1987) 231;{\bf 205B} (1988) 459;
                     {\bf 208B} (1988) 209;{\bf 231B} (1989) 65;\\
                   I. Antoniadis, G.K. Leontaris and J. Rizos, Phys. Lett.
                   {\bf 245B} (1990) 161;\\
                   G.K. Leontaris, J. Rizos and K. Tamvakis, Phys. Lett. 
                   {\bf 243B} (1990) 220;{\bf 251B} (1990) 83;\\
                   I. Antoniadis, J. Rizos and K. Tamvakis, Phys. Lett.
                   {\bf 278B} (1992) 257;{\bf 279B} (1992) 281;\\
                   J.L. Lopez and D.V. Nanopoulos, Nucl. Phys. {\bf B338}
                   (1990) 73; Phys. Lett. {\bf 251B} (1990) 73;\\
                   D. Bailin and A. Love, Phys. Lett. {\bf 280B} (1992) 26.
\bibitem{valle}   S. Ranfone and J.W.F. Valle, Phys.Lett.{\bf  B386} (1996) 151.
\bibitem{TGUT}    N. Maekawa, Prog. Theor. Phys. {\bf 106}, 401 (2001)
                  ; KUNS-1740(Proceeding, hep-ph/0110276); 
                  Phys. Lett. {\bf B521}, 42  (2001).
\bibitem{MY}      N. Maekawa and T. Yamashita, Prog. Theor. Phys. 
                  {\bf 107}, 1201 (2002);hep-ph/0303207.
\bibitem{Unif} N. Maekawa, Prog. Theor. Phys. {\bf 107}, 597 (2002);
                   N. Maekawa and T. Yamashita, Prog. Theor. Phys. 
                   {\bf 108},
                   719 (2002);Phys. Rev. Lett. {\bf 90}, 121801 (2003).
\bibitem{FN}      C.D. Froggatt and H.B. Nielsen,
                  Nucl. Phys. {\bf B147}, 277 (1979);
                  L. Ib\'a\~nez and G.G. Ross,
                  Phys. Lett. {\bf B332} (1994) 100.
\end{thebibliography}
\end{document}